\documentclass[11pt]{article}

\usepackage{palatino}
\usepackage{mathpazo}
\usepackage{braket}
\usepackage{amsfonts}
\usepackage{amssymb}
\usepackage{amsmath}
\usepackage{latexsym}
\usepackage{amsthm}
\usepackage[usenames]{color}
\usepackage{hyperref}
\usepackage{gastex}
\usepackage{float}

\hypersetup{pdfpagemode=UseNone}

\newtheorem{theorem}{Theorem}[section]
\newtheorem{corollary}[theorem]{Corollary}
\newtheorem{protocol}[theorem]{Protocol}
\newtheorem{definition}[theorem]{Definition}

\newcommand{\zo}{ \{ 0, 1 \} }
\newcommand{\eps}{\varepsilon}
\newcommand{\LTOT}{\mathrm{LT \textup{-} OT}}
\newcommand{\bLTOT}{\mathrm{\mathbf{LT \textup{-} OT}}}
\newcommand{\LTROT}{\mathrm{LT \textup{-} Random \textup{-} OT}}
\newcommand{\bLTROT}{\mathrm{\mathbf{LT \textup{-} Random \textup{-} OT}}}

\newcommand{\LTWCF}{\mathrm{LT \textup{-} WCF}}

\newcommand{\LTSCF}{\mathrm{LT \textup{-} SCF}}
\newcommand{\OT}{\mathrm{OT}}
\newcommand{\bOT}{\mathrm{\mathbf{OT}}}
\newcommand{\ROT}{\mathrm{Random \textup{-} OT}}
\newcommand{\RROT}{\mathrm{ROT}}
\newcommand{\bROT}{\mathrm{\mathbf{Random \textup{-} OT}}}
\newcommand{\WCF}{\mathrm{WCF}}
\newcommand{\bWCF}{\mathrm{\mathbf{WCF}}}
\newcommand{\SCF}{\mathrm{SCF}}
\newcommand{\bSCF}{\mathrm{\mathbf{SCF}}}

\def\I{\mathbb{1}}

\begin{document}

%-----------------------------------------------------------------------------%
\title{\bf On the existence of loss-tolerant quantum oblivious transfer protocols}
%-----------------------------------------------------------------------------%

\author{%
  Jamie Sikora\thanks{%
Department of Combinatorics \& Optimization
and Institute for Quantum Computing, University of Waterloo. Address: 200 University Ave.\ W., Waterloo, ON, N2L 3G1, Canada.
Email: {
%\tt 
jwjsikor@uwaterloo.ca}.}
}

\date{October 3, 2011}

\maketitle

\vspace{-5mm}

\begin{abstract}

Oblivious transfer is the cryptographic primitive where Alice sends one of two bits to Bob but is oblivious to the bit received. Using quantum communication, we can build oblivious transfer protocols with security provably better than any protocol built using classical communication. However, with imperfect apparatus one needs to consider other attacks. In this paper we present an oblivious transfer protocol which is impervious to lost messages.
\end{abstract}

%-----------------------------------------------------------------------------%
\section{Introduction}	        
\vspace*{-0.5pt}

%%%%%%%%%%%%%%%%%%%%%%%%%%%%%%%%

\noindent
Quantum information allows us to perform certain cryptographic tasks which are impossible using classical information alone. In 1984, Bennett and Brassard gave a quantum key distribution scheme which is unconditionally secure against an eavesdropper~\cite{M01, LC99, PS00}. This led to many new problems including finding quantum protocols for other cryptographic primitives such as \emph{coin-flipping} and \emph{oblivious transfer}.

Coin-flipping is the cryptographic primitive where Alice and Bob generate a random bit over a communication channel. We discuss two kinds of coin-flipping protocols, \emph{weak coin-flipping} where Alice wants outcome $0$ and Bob wants outcome $1$, and \emph{strong coin-flipping} where there are no assumptions on desired outcomes. We define weak coin-flipping below.

\begin{definition}[\textbf{Weak coin-flipping} ($\bWCF$)]

A \emph{weak coin-flipping} protocol, denoted $\WCF$, with cheating probabilities $(A_{\WCF}, B_{\WCF})$ and bias $\eps_{\WCF}$ is a protocol with no inputs and output $c \in \zo$ satisfying:
\begin{itemize}
\item if Alice and Bob are honest, they output the same randomly generated bit $c$;
\item $A_{\WCF}$ is the maximum probability dishonest Alice can force honest Bob to accept the outcome $c=0$;
\item $B_{\WCF}$ is the maximum probability dishonest Bob can force honest Alice to accept the outcome $c=1$;
\item $\eps_{\WCF} := \max \{ A_{\WCF}, B_{\WCF} \} - 1/2$.
\end{itemize}
\end{definition}
The idea is to design protocols which protect honest parties from cheating parties and there are no security guarantees when both parties are dishonest. 
We can assume neither party aborts in a $\WCF$ protocol. If, for instance, Alice detects Bob  has cheated then she may declare herself the winner, i.e., the outcome is $c = 0$. This is not the case in strong coin-flipping since there is no sense of ``winning.'' 

\begin{definition}[\textbf{Strong coin-flipping} ($\bSCF$)]
A \emph{strong coin-flipping} protocol, denoted $\SCF$, with cheating probabilities $(A_{\SCF}, B_{\SCF})$ and bias $\eps_{\SCF}$ is a protocol with no inputs and output $c \in \set{0, 1, \textup{abort}}$ satisfying:
\begin{itemize}
\item if Alice and Bob are honest, then they never abort and they output the same randomly generated bit $c \in \zo$;
\item $A_{\SCF}$ is the maximum probability dishonest Alice can force honest Bob to accept some outcome $c=a$, over both choices of $a \in \zo$;
\item $B_{\SCF}$ is the maximum probability dishonest Bob can force honest Alice to accept some outcome $c=b$, over both choices of $b \in \zo$;
\item $\eps_{\SCF} := \max \{ A_{\SCF}, B_{\SCF} \} - 1/2$.
\end{itemize}
\end{definition}
We note here that $\SCF$ protocols can be used as $\WCF$ protocols. The only issue is if the outcome is ``\textup{abort}''. In this case, the party who detected the cheating announces themselves the winner. Doing this, the bias in the $\WCF$ protocol is the same as in the $\SCF$ protocol.

% Strong Coin-Flipping
Aharonov, Ta-Shma, Vazirani, and Yao \cite{ATVY00} first showed the existence of an $\SCF$ protocol with bias $\eps_{\SCF} < 1/2$ followed shortly by Ambainis \cite{Amb01} who showed an $\SCF$ protocol with bias $\eps_{\SCF} = 1/4$. As for lower bounds, Mayers \cite{May97}, Lo, and Chau \cite{LC97} showed that bias $\eps_{\SCF} = 0$ is impossible.  Kitaev \cite{Kit03}, and later Gutoski and Watrous~\cite{GW07}, extended this result to show that the bias of \emph{any} $\SCF$ protocol satisfies $\eps_{\SCF} \geq 1/\sqrt{2} - 1/2$. This bound was proven to be tight  by Chailloux and Kerenidis \cite{CK09} who showed the existence of protocols with bias $\eps_{\SCF}~<~1/\sqrt{2}~-~1/2 + \delta$ for any $\delta > 0$.

% Weak Coin-Flipping
As for $\WCF$ protocols, it was shown that the bias could be less than Kitaev's bound. For example, the protocols in \cite{SR02, KN04, Moc05} provide biases of $\eps_{\WCF} = 1/\sqrt{2} - 1/2$, $\eps_{\WCF} = 0.239$, and $\eps_{\WCF} = 1/6$, respectively. The best known lower bound for $\WCF$ is by Ambainis \cite{Amb01} who showed that a protocol with bias $\eps_{\WCF}$ must use $\Omega (\log \log (1/\eps_{\WCF}))$ rounds of communication. Then, in a breakthrough result, Mochon \cite{Moc07} showed the existence of $\WCF$ protocols with bias $\eps_{\WCF} < \delta$ for any $\delta > 0$.

% Oblivious Transfer
Oblivious transfer is the cryptographic primitive where Alice sends to Bob one of two bits but is oblivious to the bit received. We define oblivious transfer and its notions of cheating below.

\begin{definition}[\textbf{Oblivious transfer} ($\bOT$)]
An \emph{oblivious transfer} protocol, denoted $\OT$, with cheating probabilities $(A_{\OT}, B_{\OT})$ and bias $\eps_{\OT}$ is a protocol \emph{with inputs} satisfying:
\begin{itemize}
\item Alice inputs two bits $(x_0, x_1)$ and Bob inputs an index $b \in \zo$;
\item when Alice and Bob are honest they never abort, Bob learns $x_b$ perfectly, Bob gets no information about $x_{\bar{b}}$, and Alice gets no information about $b$;
\item $A_{\OT}$ is the maximum probability dishonest Alice can learn $b$ without Bob aborting the protocol;
\item $B_{\OT}$ is the maximum probability dishonest Bob can learn $x_0 \oplus x_1$ without Alice aborting the protocol;
\item $\eps_{\OT} = \max \{ A_{\OT}, B_{\OT} \} - 1/2$.
\end{itemize}
\end{definition}
When a party cheats, we only refer to the probability which they can learn the desired values without the other party aborting. For example, when Bob cheats, we do not require that he learns either bit with probability $1$.

In the $\OT$ definition above there can be different ways to interpret the bias. For example, we could consider worst-case choices over inputs, we could assume the inputs are chosen randomly, etc. The protocol construction given in this paper is independent of how the inputs are chosen so this is not an issue.

Like weak coin-flipping, oblivious transfer has a related primitive which is  useful for the analysis in this paper.

\begin{definition}[\textbf{Randomized oblivious transfer} ($\bROT$)]
A \emph{randomized oblivious transfer} protocol, denoted $\ROT$, with cheating probabilities $(A_{\RROT}, B_{\RROT})$ and bias $\eps_{\RROT}$ is a protocol with \emph{no inputs} satisfying:
\begin{itemize}
\item Alice outputs two randomly generated bits $(x_0, x_1)$ and Bob outputs two bits $(b, x_b)$ where $b \in \zo$ is independently, randomly generated;
\item when Alice and Bob are honest they never abort, Bob gets no information about $x_{\bar{b}}$, and Alice gets no information about $b$;
\item $A_{\RROT}$ is the maximum probability dishonest Alice can learn $b$ without Bob aborting the protocol;
\item $B_{\RROT}$ is the maximum probability dishonest Bob can learn $x_0 \oplus x_1$ without Alice aborting the protocol;
\item $\eps_{\RROT} = \max \{ A_{\RROT}, B_{\RROT} \} - 1/2$.
\end{itemize}
\end{definition}
We note here that a protocol is considered \emph{fair} if the cheating probabilities for Alice and Bob are equal and \emph{unfair} otherwise. 

$\OT$ is an interesting primitive since it can be used to construct secure two-party protocols \cite{EGL82}, \cite{C87}, \cite{R81}. 
It was shown by Lo $\cite{Lo97}$ that $\eps_{\OT} = 0$ is impossible. This result was improved by Chailloux, Kerenidis, and Sikora \cite{CKS10} who showed that every $\OT$ protocol satisfies $\eps_{\OT} \geq 0.0586$.

% Other OT settings
Various settings for oblivious transfer have been studied before such as the bounded-storage model~\cite{DFSS08} and the noisy-storage model~\cite{S10}. In this paper, we study only information theoretic security but we allow the possibility of lost messages (more on this below). Oblivious transfer has a rich history, has various definitions, and has many names such as the \emph{set membership problem} \cite{JRS02} or \emph{private database querying} \cite{JSGBBWZ10}. 

A \emph{loss-tolerant protocol} is a quantum cryptographic protocol which is impervious to lost messages. That is, neither Alice nor Bob can cheat more by declaring that a message was lost (even if it was received) or by sending blank messages deliberately. We prefix a protocol with ``$\mathrm{LT}$-'' to indicate that it is loss-tolerant.

% LTCF
The idea of loss-tolerance was first applied to strong coin-flipping by Berlin, Brassard, Bussieres, and Godbout in \cite{BBBG08}. They showed a vulnerability in the best known coin-flipping protocol construction by Ambainis \cite{Amb01}. They circumvented this problem and presented an $\LTSCF$ protocol with bias $\eps_{\SCF} = 0.4$. Aharon, Massar, and Silman generalized this protocol to a family of $\LTSCF$ protocols with bias slightly smaller at the cost of using more qubits in the communication \cite{AMS10}. Chailloux added an encryption step to the protocol in \cite{BBBG08} to improve the bias to $\eps_{\SCF} = 0.359$ \cite{C10}. The best known protocol for $\LTSCF$ is by Ma, Guo, Yang, Li, and Wen \cite{MGYLW11} who use an EPR-based protocol which attains a bias of $\eps_{\SCF} = 0.3536$. It remains an open problem to find the best possible biases for $\LTWCF$ and $\LTSCF$. 
In fact, we do not even know if there is an $\LTWCF$ protocol with  bias less than the best possible bias for $\LTSCF$; they may in fact share the same smallest possible bias.

%LTOT
The first approach to designing loss-tolerant oblivious transfer protocols was by Jakobi, Simon, Gisin, Bancal, Branciard, Walenta, and Zbinden \cite{JSGBBWZ10}. They designed a loss-tolerant protocol for private database querying which is also known as ``$1$-out-of-$N$ oblivious transfer.'' The protocol is not technically an oblivious transfer protocol (using the definition in this paper) since an honest Bob may receive too much information. However, it is practical in the sense that it is secure against the most evident attacks. The backbone of their protocol is the use of a quantum key distribution scheme. This differs from the loss-tolerant protocol in this paper which is based on weak coin-flipping.

%%%%%%%%%%%%%%%%%%%%%%%%%%%%

\subsection*{The results of this paper}

%%%%%%%%%%%%%%%%%%%%%%%%%%%%

\noindent
We first present a protocol in Section~\ref{example} and prove it is  not loss-tolerant. Then, in Section~\ref{construction}, we show how to build $\LTOT$ protocols from $\LTWCF$ and $\LTROT$ protocols. Namely, we prove the following theorem.

\begin{theorem} \label{theorem}
Suppose there exists an $\LTWCF$ protocol with cheating probabilities $(A_{\WCF}, B_{\WCF})$ and bias $\eps_{\WCF}$ and an $\LTROT$ protocol with cheating probabilities $(A_{\RROT}, B_{\RROT})$ and bias $\eps_{\RROT}$. Then there exists an $\LTOT$ protocol with cheating probabilities
\begin{eqnarray*}
A_{\OT} & = & A_{\WCF} \, | A_{\RROT} - B_{\RROT} | + \min \{ A_{\RROT}, B_{\RROT} \}, \\
B_{\OT} & = & B_{\WCF} \, | A_{\RROT} - B_{\RROT} | + \min \{ A_{\RROT}, B_{\RROT} \}.
\end{eqnarray*}
This protocol has bias 
\[ \eps_{\OT} \leq | A_{\RROT} - B_{\RROT} | + \min \{ A_{\RROT}, B_{\RROT} \} - 1/2
= \eps_{\RROT}. \]
We have $\eps_{\OT} < \eps_{\RROT}$ when $\eps_{\WCF} < 1/2$ and $A_{\RROT} \neq B_{\RROT}$. Furthermore, the $\OT$ protocol is fair when the $\LTWCF$ protocol is fair.
\end{theorem}

In Subsection~\ref{unfair}, we show the existence of an unfair $\LTROT$ protocol with cheating probabilities $(A_{\RROT}, B_{\RROT}) = (1, 1/2)$. Combining this with the fact that there is a fair $\LTWCF$ protocol with bias $\eps_{\WCF} = 0.3536$ \cite{MGYLW11} we get the following corollary.

\begin{corollary} \label{cor}
There exists a fair $\LTOT$ protocol with bias $\eps_{\OT} = 0.4268$. 
\end{corollary}

%%%%%%%%%%%%%%%%%%%%%%%%%%%%%%%%%%%%%%%%%%%%%%%%

\section{An example of a $\bROT$ protocol that is not loss-tolerant}
\label{example}

%%%%%%%%%%%%%%%%%%%%%%%%%%%%%%%%%%%%%%%%%%%%%%%%

\noindent
In this section, we examine a protocol for $\ROT$ and show it is not loss-tolerant. This protocol has the same vulnerability as the best known coin-flipping protocol constructions based on bit-commitment, see \cite{BBBG08} for details.

\noindent
\begin{protocol}[\textbf{A} $\bROT$ \textbf{protocol~\cite{CKS10})}]
\quad
\begin{itemize} 
\item[\textup{(i)}] Bob randomly chooses $b \in \zo$ and sends Alice half of the two-qutrit state 
\noindent
\begin{equation*}
\ket{\phi_b} := \frac{1}{\sqrt{2}} \ket{bb} + \frac{1}{\sqrt{2}} \ket{22}.
\end{equation*}
\item[\textup{(ii)}] Alice randomly chooses $x_0, x_1 \in \zo$ and applies the following unitary to the qutrit 
\begin{equation*}
\ket{0} \to (-1)^{x_0} \ket{0}, \quad \ket{1} \to (-1)^{x_1} \ket{1}, \quad \ket{2} \to \ket{2}.
\end{equation*}
\item[\textup{(iii)}] Alice returns the qutrit to Bob. Bob now has the two-qutrit state 
\begin{equation*} 
\frac{(-1)^{x_b}}{\sqrt{2}} \ket{bb} + \frac{1}{\sqrt{2}} \ket{22}.
\end{equation*}
\item[\textup{(iv)}] Bob performs the measurement $\{ \Pi_0 := | \phi_b \rangle \langle \phi_b |, \; \Pi_1 := \I - \Pi_0 \}$ on the state.
\item[\textup{(v)}] If the outcome is $\Pi_0$ then $x_b=0$. If the outcome is $\Pi_1$ then $x_b=1$.
\item[\textup{(vi)}] Any lost messages are declared and the protocol is restarted from the beginning.
\end{itemize}
\end{protocol}

It has been shown in \cite{CKS10} that Bob can learn $x_0 \oplus x_1$ with probability $1$ and Alice can learn $b$ with maximum probability $3/4$. However, this does not take into account ``lost-message strategies.'' We now show such a strategy and how Alice can learn $b$ perfectly. Suppose Alice measures the first message in the computational basis. If she sees outcome ``$0$'' or ``$1$'' then she knows Bob's index $b$ with certainty. If the outcome is ``$2$'' then she replies to Bob, ``Sorry, your message was lost.'' Then they restart the protocol and Alice can measure again. Eventually, Alice will learn $b$ perfectly proving this protocol is not loss-tolerant.

This protocol illustrates another interesting point about the design of $\OT$ protocols. One may not be able to simply change the amplitudes in the starting states to balance the cheating probabilities. For example, if we were to change the amplitudes in $\ket{\phi_b}$, then Bob would have a nonzero probability of getting the wrong value for $x_b$. Thus, balancing an unfair $\OT$ protocol is not as straightforward as it can be in coin-flipping.

%%%%%%%%%%%%%%%%%%%%%%%%%%%%%%%%%%%%%%%%%%%%%%%%

\section{Constructing loss-tolerant oblivious transfer protocols}
\label{construction}

%%%%%%%%%%%%%%%%%%%%%%%%%%%%%%%

\noindent
In this section, we prove Theorem~\ref{theorem} by constructing an $\LTOT$ protocol from an $\LTWCF$ protocol and a (possibly unfair) $\LTROT$ protocol. In doing so, we have to overcome some issues that are not present when designing $\LTSCF$ protocols. These issues include:
\begin{itemize}
\item it is not always possible to simply reset a protocol with inputs;
\item balancing the cheating probabilities can be difficult;
\item it is not possible to switch the roles of Alice and Bob since Bob must be the receiver;
\item an honest party must not learn extra information about the other party's inputs (or outputs in the case of $\ROT$).
\end{itemize}
We deal with these issues by reducing the problem one step at a time. First we reduce the task of finding $\LTOT$ protocols to finding $\LTROT$ protocols in Subsection~\ref{same}. Then we build an $\LTROT$ protocol from an $\LTWCF$ protocol and two (possibly unfair) $\LTROT$ protocols in Subsection~\ref{wcf}. In Subsection~\ref{symmetry}, we show how to create the two $\LTROT$ protocols from a single $\LTROT$ protocol. Finally, we show an unfair $\LTROT$ protocol in Subsection~\ref{unfair} to prove Corollary~\ref{cor}.

%%%%%%%%%%%%%%%%%%%%%%%%%%%%%%%

\subsection{Equivalence between $\bLTOT$ protocols and $\bLTROT$ protocols with respect to bias}
\label{same}

%%%%%%%%%%%%%%%%%%%%%%%%%%%%%%%

\noindent
Having a protocol with inputs is an issue when building protocols loss-tolerantly. In recent $\LTSCF$ protocols, if messages were lost for any reason, then the protocol is simply restarted at some point, but this is not always an option with $\OT$ because the inputs could have context, e.g., Alice's bits could be database entries. For this reason, we cannot simply ``reset'' them and repeat the protocol. To remedy this issue, we use $\ROT$.

It is well known that $\OT$ and $\ROT$ share the same cheating probabilities, i.e., if there exists an $\OT$ protocol with cheating probabilities $(A_{\OT}, B_{\OT}) = (x,y)$ then there exists a $\ROT$ protocol with cheating probabilities $(A_{\RROT}, B_{\RROT}) = (x,y)$, and vice versa. For completeness, we show these reductions and prove  they preserve loss-tolerance.

\begin{protocol}[$\bLTROT$ \textbf{from} $\bLTOT$]
\quad 
\begin{itemize}
\item[\textup{(i)}] Alice randomly chooses $x_0, x_1 \in \zo$ and Bob randomly chooses $b \in \zo$.
\item[\textup{(ii)}] Alice and Bob input the choices of bits above into the $\LTOT$ protocol so that Bob learns $x_b$.
\item[\textup{(iii)}] Alice outputs $(x_0, x_1)$ and Bob outputs $(b, x_b)$.
\end{itemize}
\end{protocol}

It is straightforward to see that this reduction preserves the loss-tolerance of the $\LTOT$ protocol since we are only restricting how the inputs are chosen. More interesting is the reduction from $\LTROT$ to $\LTOT$.

\begin{protocol}[$\bLTOT$ \textbf{from} $\bLTROT$]
\quad 
\begin{itemize}
\item[\textup{(i)}] Alice and Bob decide on their desired choices of inputs to the $\LTOT$ protocol.
\item[\textup{(ii)}] Alice and Bob use an $\LTROT$ protocol to generate the output $(x_0, x_1)$ for Alice and $(b, x_b)$ for Bob.
\item[\textup{(iii)}] Bob tells Alice if his output bit $b$ is equal to his desired index. If it is not equal, Bob changes it and Alice switches her two bits.
\item[\textup{(iv)}] Alice tells Bob which of her two bits $(x_0, x_1)$ are equal to her desired inputs. Alice and Bob flip their outcome bits accordingly. \end{itemize}
\end{protocol}

This reduction is a way to derandomize the outputs of the $\LTROT$ protocol.
We see that this also preserves the loss-tolerance of the $\LTROT$ protocol since classical information can simply be resent if lost in transmission.

Using the reductions above, we have reduced the task of finding $\LTOT$ protocols to finding $\LTROT$ protocols.

%%%%%%%%%%%%%%%%%%%%%%%%%%%%%%%

\subsection{Creating $\bLTROT$ protocols}
\label{wcf}

%%%%%%%%%%%%%%%%%%%%%%%%%%%%%%%

\noindent
There is a simple construction of an $\SCF$ protocol with bias $\eps \approx 3/4$ and it proceeds as follows. Alice and Bob first use a $\WCF$ protocol with bias $\eps \approx 0$. The ``winner'' gets to flip a coin to determine the outcome of the $\SCF$ protocol. Of course, a dishonest player would like to ``win'' the $\WCF$ protocol since then they have total control of the $\SCF$ outcome.

We mimic this idea to create a protocol prototype for $\LTROT$ and discuss why it does not work.

\noindent
\begin{protocol}[\textbf{A protocol prototype}]
\quad
\begin{itemize}
\item[\textup{(i)}] Alice randomly chooses two bits $(x_0, x_1)$ and Bob randomly chooses an index $b \in \zo$.
\item[\textup{(ii)}] Alice and Bob perform an $\LTWCF$ protocol with bias $\eps_{\WCF}$ to create random $c \in \zo$.
\item[\textup{(iii)}] If $c = 0$, then Bob sends $b$ to Alice. Alice then replies with $x_b$.
\item[\textup{(iv)}] If $c = 1$, then Alice sends $(x_0, x_1)$ to Bob.
\end{itemize}
\end{protocol}
This protocol has bias $\eps_{\RROT} < 1/2$ if $\eps_{\WCF} < 1/2$. However, the problem is that honest Alice learns $b$ with probability $3/4$ when Bob is honest. This is simply not allowed in a $\ROT$ protocol because honest Alice should never obtain any information about $b$. Honest Bob learns $x_0 \oplus x_1$ with probability  $3/4$, which is also not allowed since he should only learn $x_0$ or $x_1$. This illustrates another issue when designing $\OT$ and $\ROT$ protocols.

To remedy this problem, instead of Alice and Bob revealing their bits entirely, they can use (possibly unfair) $\LTROT$ protocols. We present a modified version of the protocol below.

\noindent
\begin{protocol}[\textbf{An} $\bLTROT$ \textbf{protocol}]
\quad
\begin{itemize}
\item[\textup{(i)}] Alice and Bob perform an $\LTWCF$ protocol with cheating probabilities $(A_{\WCF}, B_{\WCF})$ and bias $\eps_{\WCF}$ to create random $c \in \zo$.
\item[\textup{(ii)}] If $c = 0$, then Alice and Bob generate their outputs using an $\LTROT$ protocol with cheating probabilities $(A_{\RROT}, B_{\RROT}) = (x,y)$, where $x \geq y$.
\item[\textup{(iii)}] If $c = 1$, then Alice and Bob generate their outputs using an $\LTROT$ protocol with cheating probabilities $(A_{\RROT}, B_{\RROT}) = (y,x)$.
\item[\textup{(iv)}] Alice and Bob abort if and only if either $\LTROT$ protocol is aborted.
\end{itemize}
\end{protocol}

We now prove that this $\LTROT$ protocol has cheating probabilities equal to those in Theorem~\ref{theorem}. We show it for cheating Alice as the case for cheating Bob is almost identical. Since $x \geq y$, Alice would prefer the outcome of the $\WCF$ protocol to be $c=0$. She can force $c=0$ with probability $A_{\WCF}$ and in this case she can learn $b$ with probability $x$. If $c=1$, she can learn $b$ with probability $y$. Letting $A'_{\RROT}$ be the amount she can learn $b$ in the protocol above, we have 
\[ A'_{\RROT} = A_{\WCF} \, x + (1-A_{\WCF}) \, y = A_{\WCF} \, (x-y) + y. 
\]

All that remains to prove Theorem~\ref{theorem} is to show that an $\LTROT$ protocol with cheating probabilities $(A_{\RROT}, B_{\RROT}) = (\alpha, \beta)$ implies the existence of an $\LTROT$ protocol with cheating probabilities $(A_{\RROT}, B_{\RROT}) = (\beta, \alpha)$, for any $\alpha, \beta \in [1/2, 1]$. This way, we can just set $x = \max \{ \alpha, \beta \}$ and $y = \min \{ \alpha, \beta \}$.

%%%%%%%%%%%%%%%%%%%%%%%%%%%%%%

\subsection{Symmetry in $\bLTROT$ protocols}
\label{symmetry}

%%%%%%%%%%%%%%%%%%%%%%%%%%%%%%

\noindent
Suppose we have an $\LTROT$ protocol with cheating probabilities $(A_{\RROT}, B_{\RROT}) = (\alpha, \beta)$, for some $\alpha, \beta \in [1/2, 1]$. We now show how to create an $\LTROT$ protocol with cheating probabilities $(A_{\RROT}, B_{\RROT}) = (\beta, \alpha)$. The trick is to switch the roles of Alice and Bob.

\noindent
{\begin{protocol}[\textbf{A} $\bROT$ \textbf{protocol (randomized version of a protocol in~\cite{WW06})}]
\quad
\begin{enumerate}
\item[\textup{(i)}] Alice and Bob use an $\LTROT$ protocol with cheating probabilities $(A_{\RROT}, B_{\RROT}) = (\alpha, \beta)$ except that Bob is the sender and Alice is the receiver. Let Alice's output be $(b, x_{b})$ and let Bob's output be $(x_0, x_1)$.
\item[\textup{(ii)}] Alice randomly chooses $d \in \zo$ and sends $d \oplus x_{b}$ to Bob.
\item[\textup{(iii)}] Alice outputs $(x'_0, x'_1) = (d, d \oplus b)$ and Bob outputs $(b', m) = (x_0 \oplus x_1, d \oplus x_{b} \oplus x_0)$.
\item[\textup{(iv)}] Alice and Bob abort if and only if the $\LTROT$ protocol is aborted.
\end{enumerate}
\end{protocol}}
Notice this protocol is loss-tolerant since classical messages can be resent if lost in transmission. We can write Bob's output $m$ as $d \oplus x_{b} \oplus x_0 = d \oplus bb'$. Thus, if $b'=0$ then $m = d = x'_0$ and if $b' = 1$ then $m = d \oplus b = x'_1$. Therefore Bob gets the correct value for $x'_{b'}$. Since $x'_0 \oplus x'_1 = d \oplus (d \oplus b) = b$, honest Bob gets no information about Alice's other bit and cheating Bob can learn $x'_0 \oplus x'_1$ with maximum probability $\alpha$. Since $b' = x_0 \oplus x_1$, honest Alice gets no information about $b'$ and cheating Alice can learn $b'$ with maximum probability $\beta$. Therefore, $(A_{\RROT}, B_{\RROT}) = (\beta, \alpha)$ as desired. Since $b, x_0, x_1$, and $d$ are all randomly generated, so are $x'_0, x'_1$, and $b'$ making this a valid $\LTROT$ protocol.

This completes the proof of Theorem~\ref{theorem}.

%%%%%%%%%%%%%%%%%%%%%%%%%%%%%%%%%%%%%%%%%%%%%%%%%

\subsection{An unfair $\bLTROT$ protocol} 
\label{unfair}

%%%%%%%%%%%%%%%%%%%%%%%%%%%%%%%%%%%%%%%%%%%%%%%%%

\noindent
We present here an $\LTROT$ protocol with cheating probabilities $(A_{\RROT}, B_{\RROT}) = (1/2, 1)$. Note that even though this protocol has bias $\eps_{\RROT} = 1/2$, it can be used to create a protocol with smaller bias using recent $\LTWCF$ protocols and Theorem~\ref{theorem}.

\noindent
\begin{protocol}[\textbf{An unfair} $\bLTROT$ \textbf{protocol}]
\quad
\begin{itemize}
\item[\textup{(i)}] Bob randomly chooses an index $b \in \zo$ and another random bit $d \in \zo$.
\item[\textup{(ii)}] Bob sends Alice the qubit $H^b \ket{d}$.
\item[\textup{(iii)}] Alice randomly chooses $x_0, x_1 \in \zo$ and applies the unitary $X^{x_0} Z^{x_1}$ to the qubit.
\item[\textup{(iv)}] Alice returns the qubit to Bob which is in the state $X^{x_0} Z^{x_1} H^b \ket{d} = H^b \ket{x_b \oplus d}$ (up to global phase).
\item[\textup{(v)}] Bob has a two-outcome measurement (depending on $b$ and $d$) to learn $x_b$ perfectly.
\item[\textup{(vi)}] If any messages are lost the protocol is restarted from the beginning.
\end{itemize}
\end{protocol}
We see that this is a valid $\ROT$ protocol. Firstly, because honest Bob learns $x_b$ and gets no information about $x_{\bar b}$ (since $H^b \ket{x_b \oplus d}$ does not involve $x_{\bar{b}}$). Secondly, Alice cannot learn any information about $b$, even if she is dishonest, since the density matrices for $b=0$ and $b=1$ are identical. Therefore $A_{\RROT} = 1/2$. This protocol is loss-tolerant concerning cheating Alice since $b$ and $d$ are reset if any messages are lost so Alice cannot accumulate useful information. It is also loss-tolerant concerning cheating Bob since he can already learn both of Alice's bits perfectly. He can do this by first sending Alice half of 
\begin{equation*}
\ket{\Phi^+} = \dfrac{1}{\sqrt 2} \ket{00} + \dfrac{1}{\sqrt 2} \ket{11}.
\end{equation*}
Each choice of $(x_0, x_1)$ corresponds to Bob having a different Bell state at the end of the protocol. From this, $x_0$ and $x_1$ can be perfectly inferred, yielding $B_{\RROT} = 1$.

%%%%%%%%%%%%%%%%%%%%%%%%%%%%%%

\section{Conclusions and open questions}

%%%%%%%%%%%%%%%%%%%%%%%%%%%%%%

\noindent
We have designed a way to build $\LTOT$ protocols by using an $\LTWCF$ protocol to help balance the cheating probabilities in a (possibly unfair) $\LTROT$ protocol. This protocol uses well known reductions between $\OT$ and $\ROT$ and the reduction to switch the roles of Alice and Bob.

The construction in this paper is robust enough to design $\OT$ protocols with other definitions of cheating Bob. Suppose that Bob  wishes to learn $f(x_0, x_1)$ where $f \neq \mathrm{XOR}$ is some functionality. In this case, we may not be able to switch the roles of Alice and Bob in a way that switches the cheating probabilities as in Subsection~\ref{symmetry}. However, instead of just using one $\LTROT$ protocol and creating another from it, we could have just as easily used two different $\LTROT$ protocols (with a consistent notion of cheating Bob).

A limitation of this protocol design is that is uses $\LTROT$ protocols as subroutines. Even if $\LTWCF$ protocols with bias $\eps_{\WCF} \approx 0$ are constructed, using the protocols in Subsection~\ref{unfair} can reduce the bias to only $\eps_{\OT} \approx 1/4$. It would be interesting to see if there exists an $\LTOT$ protocol with cheating probabilities $(A_{\OT}, B_{\OT}) = (\alpha, \beta)$ where $\alpha + \beta < 3/2$.

An open question is to show if using more $\LTWCF$ subroutines can help improve the bias. In \cite{CK09}, many $\WCF$ protocols were used to drive the bias of a $\SCF$ protocol down towards the optimal value of $1/\sqrt{2} - 1/2$. Can something similar be done for $\OT$ or $\LTOT$?

%%%%%%%%%%%%%%%%%%%%%%%%%%%%%%

\section*{Acknowledgements}

%%%%%%%%%%%%%%%%%%%%%%%%%%%%%%

\noindent
I would like to thank Ashwin Nayak and Levent Tun\c cel for helpful discussions. I acknowledge support from NSERC, MITACS, and ERA (Ontario). 

%%%%%%%%%%%%%%%%%%%%%%%%%%%%%%

\bibliographystyle{alpha}
\bibliography{paper}

\end{document}